\documentstyle[epsf,prl,aps,multicol]{revtex}
\begin{document}
\title{Metal-Insulator Transition and Ferromagnetism in Diluted Magnetic Semiconductors}
\author{S.-R. Eric Yang$^{1,2}$ and A.H.MacDonald$^2$}
\address{
1)Department of Physics, Korea University, Seoul 136-701, Korea \\
2)Department of Physics, University of Texas at Austin, Austin TX 78712 }
\maketitle
\draft
\begin{abstract}

\noindent
We have investigated the interplay between the metal-insulator 
transition and ferromagnetism in ${\rm III}_{1-x}{\rm Mn}_x{\rm V}$ 
semiconductors.  Our study is based on a model in which $S=5/2$ Mn local
moments are exchange-coupled to valence band holes that interact via Coulomb
interactions with each other, with ionized
Mn acceptors, and with the antisite defects present in these materials.   
We find quasiparticle participation ratios consistent with
a metal-insulator transition that occurs in the ferromagnetic state near  
$x \sim 0.01$.  By evaluating the distribution of mean exchange
fields at Mn moment sites, we provide evidence in favor
of the applicability of impurity-band magnetic-polaron and hole-fluid models on insulating 
and metallic sides of the phase transition respectively.  

\end{abstract}

\thispagestyle{empty}
\pacs{PACS numbers: 75.50.Pp, 75.10.-b, 75.30.Hx}
\begin{multicols}{2}

\noindent
{\it Introduction}.- Recent progress\cite{recentrefs} in the growth of 
diluted magnetic semiconductors semiconductors that exhibit
ferromagnetism\cite{classicrefs}
at relatively high temperatures has suggested exciting new possibilities 
for devices that combine information processing and storage functionalities
in a single material.  It is generally accepted\cite{ourbookchapter} that 
Mn acts as an acceptor in these semiconductors, that its half-filled 
d-shell contributes a $S=5/2$ local moment\cite{szc,lin,fed} to the system's low-energy
degrees of freedom, and that ferromagnetism is due to interactions between
local moments that are mediated by valence-band holes.
Recent experiments\cite{ohno1,mat,mitexpt,pot} demonstrate that 
ferromagnetism occurs in both metallic and insulating states,
and that both magnetic and transport properties are sensitive to 
the Mn fraction $x$ and to the density of compensating 
antisite and other defects in the material. The highest ferromagnetic
critical temperatures ($T_c$) appear\cite{pot} to occur in the most 
metallic samples.  The role of Coulomb interactions in the ferromagnetism in these 
materials is subtle.  At high carrier densities, well on the metallic side of 
the metal-insulator transition, exchange and correlation in the hole system
is expected\cite{jungwirthtc} to enhance ferromagnetism.  
Well on the insulator side of the transition, however, 
Coulomb interactions create a Mott gap that increases the importance
of randomness in Mn ion positions, suppresses 
carrier-hopping between Mn sides, and eventually 
turns off the coupling between local moments that can lead to ferromagnetism.

\begin{figure}
%\hspace{1truecm}
%\vspace{-0.5truecm}
\center
\centerline{\epsfysize=2.3in
\epsfbox{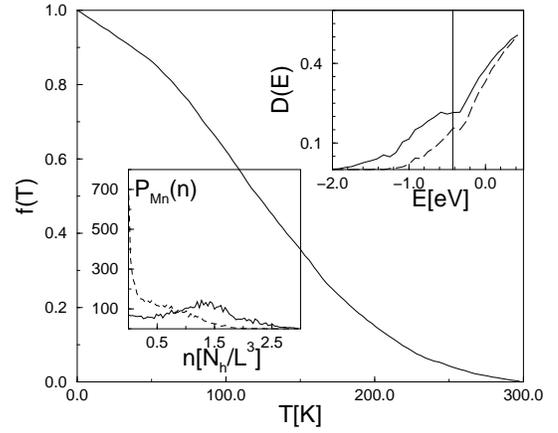}}
%\vspace{-0.5truecm}
\begin{minipage}[t]{8.1cm}
\caption{$f(T)$, the fraction of Mn that experience a mean-field stronger
than $k_BT$, {\it vs.} T in the metallic case. ($N_{Mn}=1.0 {\rm nm}^{-3}$,
$n_h=N_{Mn}/3$.) $P_{Mn}(n)$, the probability distribution function
for partial hole densities at Mn sites in units of the average
hole density, is plotted for majority (solid line) and minority (dashed line) spins in
one inset and the quasiparticle density-of-states in the other inset.
For this case the hole density at an isolated Mn is 1.63 times larger
than the average hole density.  The density of states $D(E)$
is per occupied hole with energies in eV.  Note the anomaly at
the Fermi energy $E_F$, indicated by a vertical line.
The number of disorder realization is fifteen for $D(E)$ and ten for $f(T)$ and $P_{Mn}(n)$.
A a simulation
cell of side $L=8 {\rm nm}$ is used. }
\label{figure1}
\end{minipage}
\end{figure}

\begin{figure}
%\hspace{1truecm}
%\vspace{-0.5truecm}
\center
\centerline{\epsfysize=2.3in
\epsfbox{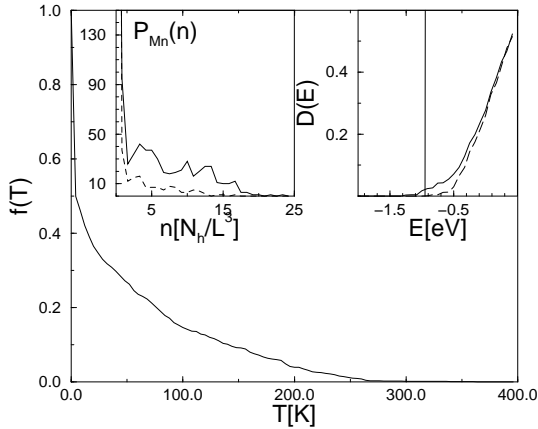}}
%\vspace{-0.5truecm}
\begin{minipage}[t]{8.1cm}
\caption{
As in Figure 1 but for the insulating case.
($N_{Mn}= 0.25 {\rm nm}^{-3}$ and $n_h=N_{Mn}/10$).
The hole density at an isolated Mn site is
$19.9$ times larger than the average hole density in this case.}
\label{figure2}
\end{minipage}
\end{figure}

In this Letter we report on a numerical Hartree-Fock study of model (III,Mn)V 
ferromagnets in both metallic and insulating regimes.  By evaluating the 
distribution of $T=0$ Mn exchange mean-fields, defined precisely below, 
we find evidence that is generally
supportive of the impurity-band magnetic-polaron\cite{bhatt,dassarma,timm1} 
picture that has been used to
describe ferromagnetism in the insulating regime, and of the hole-fluid 
picture that has been used\cite{ourbookchapter,jungwirthtc,dietl,abolfath}
in the metallic regime.  Our principle results are 
summarized in Figs.~\ref{figure1} and \ref{figure2} where we plot ${\it vs.} T$ 
the fraction of Mn sites that have exchange mean-fields larger than $k_B T$.
\noindent\\
In the metallic case\cite{compensation} ($x=0.05$ and $p = 3.3 \times 10^{20} {\rm cm}^{-3}$) 
the distribution of coupling strengths is peaked at a finite value close to
its uniform hole-fluid value and all Mn ions are strongly coupled to holes.
In the insulating case ($x=0.01$ and hole density
$p = 2.8 \times 10^{19} {\rm cm}^{-3}$), on the other hand, 
we find that that vast majority of Mn spins are nearly free  the few strongly coupled 
Mn ions will form magnetic polarons that 
grow slowly in size and interact more strongly as the temperature is lowered.
In the following sections we explain how these results were obtained 
and elaborate on their significance for models of diluted magnetic 
semiconductor ferromagnetism.

\noindent
{\it Model Hamiltonian}.- Most theoretical work on (III,Mn)V ferromagnets 
has started from one of two idealized limits.  Impurity-band models\cite{bhatt,dassarma,timm1} 
achieve simplification by assuming that the holes that mediate interactions between
Mn ion moments are strongly localized, whereas hole-fluid models achieve
simplification by neglecting disorder due to Mn acceptors and other defects and 
treating it perturbatively in estimating transport
coefficients.  An insulator to metal phase transition 
occurs in these ferromagnets as Mn and hole densities increase; 
experimentally\cite{recentrefs} it appears that the metal-insulator transition 
usually occurs near $x \sim 0.01$, 
likely depending on the density of compensating antisite defects, 
a quantity that is expected to be sensitive the details of sample-growth
and annealing protocols.  Although it is generally agreed that an impurity band picture should
apply far on the insulator side of the transition and a hole-fluid model far on the 
metallic side of the transition, it has not been clear which approach is a better
starting point in the experimentally relevant parameter ranges.  We address this issue  
by examining the ferromagnetic ground state using a
model, and an approximation scheme, that captures the physics of both limits
correctly.  Our study is built on $\vec{k}\cdot\vec{p}$ envelope-function 
approach description
of the valence band\cite{bandtheory} and a Hartree-Fock description of interactions.  

The four terms in the single-particle part of our band electron model Hamiltonian, 
$H^0=H^K+H^{K.ex.}+H^{Mn-h}+H^{as-h}$, require some discussion. 
$H^{K.ex.}$ represents the kinetic exchange interaction between Mn local moments, assumed
to be aligned in the ferromagnetic system ground state\cite{noncollinear}, 
$H^{Mn-h}$ represents the attractive Coulomb interaction between an ionized $Mn^{2+}$
acceptor and a valence band hole.  In an envelope function formalism, central cell
corrections to this interaction are necessary\cite{Bhatta}
to achieve an accurate description of the isolated bound-acceptor limit.
$H^{as-h}$ describes the repulsive interaction between holes 
and antisite defects (represented as sites with charge +2).
These defects act as deep donors partially compensating the Mn acceptors and reducing the overall 
hole density, and provide an important additional scattering center.
$H^K$ is the usual kinetic energy term.  In this study we 
ignore\cite{mat,abolfath,janko} mixing between heavy and light hole bands by using a simple 
parabolic band approximation. 
\begin{eqnarray}
&H^0&=\frac{-\hbar^2}{2m}\vec{\nabla}+
\frac{1}{2}S\sum_{I} \hat{\Omega}_I\cdot\vec{\tau}J(\vec{r}-\vec{R_I})\nonumber\\
&+&\sum_{I}(-\frac{e^2}{\epsilon|\vec{r}-\vec{R_I}|}-
V_0e^{-|\vec{r}-\vec{R_I}|^2/r_0^2})
+\sum_{K}\frac{2e^2}{\epsilon|\vec{r}-\vec{R_K}|}.\nonumber\\
\label{charge}
\end{eqnarray} 
Here $J(\vec{r})=(J_{pd}/(2\pi a_0^2)^{3/2})\, \exp(-r^2/2{a_0}^2)$, 
$\vec{\tau}=(\tau_x,\tau_y,\tau_z)$ is the Pauli spin matrix vector, 
I labels Mn sites, K labels antisite defect sites, and 
$\hat{\Omega}_I$ is the orientation of the $I^{th}$ Mn spin. 
The term in the potential proportional to $V_0$ is the central cell correction.
Both Mn ions and antisites were distributed randomly\cite{timm2} 
in a cube of side $L$.  The long range of the Coulomb interaction requires overall 
charge neutrality so that $n_h-N_{Mn}+2N_{as}=0$,
where $n_h$ is the density of holes, $N_{Mn}$ the density of Mn ions and 
$N_{as}$ the density of antisites.   
In the ground state, all Mn spins are oriented along the $\hat z$ direction and 
$H^{0}$ is block diagonal in its spin indices.  In the spin wave configurations 
discussed below, $\hat x$ and $\hat y$ components of the spins 
are present, doubling the dimension of the matrix 
that must be diagonalized numerically.  The wavevector cutoff in the quasiparticle 
wavefunction expansion\cite{hamiltoniandetails} was tested by computing 
the binding energy of a hole to an isolated Mn, comparing
with the results of Bhattacharjee and Benoit a la Guillaume\cite{Bhatta}
who find that a binding energy of 112meV, 86 meV when the kinetic exchange 
term is neglected, and 68meV when the central cell correction is also neglected.
Our results are 124meV,  88 meV, and 48meV respectively, demonstrating 
an adequate description of the completely isolated Mn limit.

In order to capture the correct physics of both metallic and impurity band limits,
hole-hole interactions must be described using an approximation
that accounts for screening effects in the metallic regime and 
avoids artificial self-interaction effects in the localized regime,
motivating our use of Hartree-Fock (HF) theory \cite{Yang1}.
The HF quasiparticle Hamiltonian is $ H^{HF}=H^0+(V^H+V^X)$ where 
the Hartree and Fock matrix elements, $V^H_{\vec{k}\sigma,\vec{k'}\sigma'}$ and
$V^X_{\vec{k}\sigma,\vec{k'}\sigma}$ , can be expressed in terms of
the density matrix.  We evaluate the density matrix in the $k$-space representation
used to expand our envelope function quasiparticle wavefunctions:
$\rho_{\vec{k}\sigma,\vec{k'}\sigma'}
=\sum_{\alpha}n_{\alpha}
c_{\vec{k}\sigma}^{(\alpha)}c_{\vec{k'}\sigma'}^{(\alpha)*}$,
where $|\alpha\rangle=\sum_{\vec{k}\sigma}c_{\vec{k}\sigma}^{(\alpha)}|
\vec{k}\sigma\rangle$,  
and $n_{\alpha}$ is a quasiparticle occupation number. 
Our HF scheme becomes {\it exact} in the strongly localized limit, since a localized 
quasiparticle does not interact with itself, but is expected to 
overestimate inhomogeneity in the metallic limit because it neglects 
quantum fluctuations in the many-hole state.
\begin{figure}
%\hspace{1truecm}
%\vspace{-0.5truecm}
\center
\centerline{\epsfysize=2.3in
\epsfbox{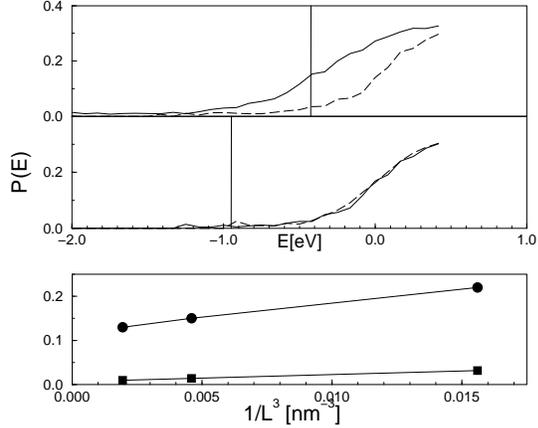}}
%\vspace{-0.5truecm}
\begin{minipage}[t]{8.1cm}
\caption{Participation ratios {\it vs.} energy in a simulation cell of side
$L=8{\rm nm}$ for majority (solid line) and minority (dashed line)
quasiparticles.  The Fermi energies are
indicated by solid vertical lines.
The upper panel is for the metallic case of Fig. 2 and the middle panel is
for the insulating case of Fig. 3 and these results were obtained by averaging
over 15 disorder realizations.
The bottom panel shows the system size dependence of majority-spin
participation ratios at the Fermi energy for the two cases.}
\label{figure3}
\end{minipage}
\end{figure}

\noindent
{\it Participation ratios}.- 
In order to verify that our model correctly describes the metal-insulator transition 
in these materials, we measured the localization properties of our Hartree-Fock 
quasiparticles by evaluating their participation ratios,
$P_{\alpha}=1/(L^3\int d^3r |\Psi_{\alpha}(\vec{r})|^4)$, 
where $\Psi_{\alpha}(\vec{r})=\langle\vec{r}|\alpha\rangle$ is a
normalized quasiparticle wavefunction.  
(For a localized state $P_{\alpha} \sim (\xi_{\alpha}/L)^3$, 
where $\xi_{\alpha}$ is the localization length.)  
Fig. 3 compares participation ratios evaluated at $ x=0.05$ 
and $0.0125$, assuming compensated carrier densities $n_h=N_{Mn}/3$ and 
$n_h=N_{Mn}/10$, respectively.
The participation fractions extrapolated to infinite volume
are clearly finite and therefore consistent with metallic transport at $x=0.05$
whereas they are consistent with zero and localized quasiparticles at $x=0.0125$,
in agreement with experiment.

\noindent
{\it Impurity-Band Picture {\it vs.}Hole-Fluid Picture}.- 
The exchange mean-fields used to construct Fig.~\ref{figure1} and Fig.~\ref{figure2} 
are given by $H_{eff,I} = \int d \vec r J(\vec r - \vec R_I) (n_{\downarrow}(\vec r) - 
n_{\uparrow}(\vec r))/2 \approx J_{pd} (n_{\downarrow}(\vec R_I)-n_{\uparrow}(\vec R_I))/2$
where the partial densities are determined by solving the Hartree-Fock equations self-consistently.
In the extreme impurity band limit, $H_{eff,I}$ would have the value $J_{pd} n_{max}/2
\approx 25 {\rm meV}$  for a fraction $n_{h}/N_{Mn}$ of the local moments
and much smaller values for all others.  Our findings are in qualitative agreement, although
the peak $H_{eff}$ value is somewhat smaller and there is no sharp peak in the distribution
function near the peak value, presumably because of the variable Mn and antisite defect environment
experienced by localized holes.
(The shallow impurity Bohr radius $a_B^*$ calculated using the heavy-hole mass is $\sim 1.0 {\rm nm}$;
the peak hole density, $4.9 \times 10^{20} {\rm cm}^{-3}$ is slightly larger than the
shallow impurity value because of central cell corrections.) 
Kaminski and Das Sarma\cite{dassarma} have recently estimated the impurity-band
magnetic polaron model $T_c$, relating it to $N_{Mn}$, $n_h$ and the 
maximum mean-field exchange coupling by $T_c  \sim 0.5 (N_{Mn}/n_h)^{1/2}
(n_h{a_B^*}^3)^{1/6} H_{max} \exp(-0.86/{n_h}^{1/3}a_B^*) /k_B$.
Naively applying this formula to our $x=0.01$ case with the shallow impurity Bohr radius
yields $T_c \sim 15K$.  Note that $T_c$ is expected to decrease rapidly at lower hole densities 
because of the weakening magnetic polaron coupling. 
These estimated $T_c$'s are somewhat larger than what has typically been observed 
at $x \sim 0.01$ in (Ga,Mn)As, possibly because the theory neglects Pauli exclusion principle 
effects, superexchange interactions, that favor opposite orientations of nearby
impurity-band holes and oppose ferromagnetism.  Nevertheless, it appears that 
this picture provides an excellent qualitative description of ferromagnetism at 
$x \sim 0.01$ and smaller.

In the metallic case, the exchange mean-field 
distribution differs qualitatively, with strong couplings common and 
few weakly coupled moments.  Our results likely overestimate the degree of 
density variation in samples with $x \sim 0.05$, both because we have neglected
Mn acceptor-antisite correlations\cite{timm2} and because of our use of the Hartree-
Fock approximation.  (The corresponding self-consistent Hartree calculations 
lead to more sharply peaked Mn site majority-spin density distributions.) 
We note that the most likely hole density at a Mn site is 
approximately 1.5 times larger than the average hole density and that the 
most likely mean-field coupling strengths are again somewhat larger than
the uniform hole-fluid value $H_{fl} = n J_{pd} /2 \approx 190$K.  

There are important differences between metallic and insulating cases in the physics
that controls $T_c$.  The metallic $T_c$ can be limited either\cite{schliemann,millis} by
the system's stiffness against collective magnetization orientation 
variation, or if the stiffness is large enough by the competition between
local moment entropy, exchange interactions, and the band energy cost of hole-spin polarization 
that is captured by the hole-fluid mean-field-theory $T_c$ expression\cite{dietl97}.
We have estimated the magnetic stiffness of both insulating and 
metallic states by evaluating the HF electronic energy cost of an imposed spin-wave
with wavevector $Q$.  If disorder were neglected, the procedure we follow
in evaluating spin spin-wave energies differs from 
the theory\cite{kon1} of K\" onig {\it et al.} only through retardation effects.
\begin{figure}
%\hspace{1truecm}
%\vspace{-0.5truecm}
\center
\centerline{\epsfysize=2.3in
\epsfbox{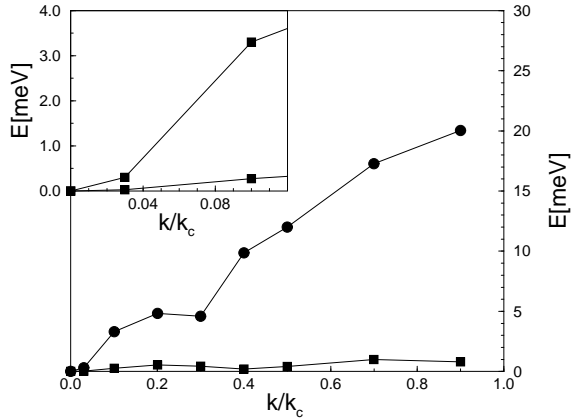}}
%\vspace{-0.5truecm}
\begin{minipage}[t]{8.1cm}
\caption{Average spin-flip energy {\it vs.} wavevector. The
upper curve is for the metallic case of Fig.1 while the lower curve is for
the insulating case of Fig.2.  The wavevector is normalized to a Debye wavevector
defined by the Mn density in each case, $k_c=(6\pi^2N_{Mn})^{1/3}$.
The inset shows a magnified view of  the small wavevector regime.}
\label{figure4}
\end{minipage}
\end{figure}
\noindent\\
Figure 4 displays average spin reversal energies for metallic and insulating states.
Note that the energy cost of slow (long wavelength) spin-direction variations is extremely
small in the insulating case because these are sensitive mainly to the weak interactions 
between magnetic polarons that are necessary for long-range magnetic order.  It follows
that mean-field will completely fail in estimating the critical temperature of insulating 
DMS ferromagnets.  The energy cost for spin-direction variations that we find
in the metallic case is in qualitative agreement with hole-fluid model results, although 
it is larger at short wavelengths.  In the hole-fluid mean-field theory the energy
cost of spin-direction variations would equal $H_{fl}$ at all wavevectors.  It follows that 
for the parabolic band model we have studied substantial corrections\cite{schliemann,millis}
would apply to its mean-field $T_c$.  It is important to note, however, that valence band spin-orbit
interactions stiffen collective spin-variations in these ferromagnets\cite{kon2}, reducing
corrections to mean-field-theory $T_c$ estimates.

SREY is supported in part by the KOSEF Quantum-functional Semiconductor Research Center at 
Dongkuk University.  Work at the University of Texas was supported by 
by the Welch foundation and by DARPA/ONR Award No. N00014-00-1-095.

\end{multicols}


\begin{references}
\bibitem{timm2} C. Timm {\it et al.} have pointed out that defect 
correlations, which we do not include here, will reduce the importance of 
disorder, expanding the validity range of the hole-fluid picture.
\bibitem{hamiltoniandetails} 
The model Hamiltonian parameters we use are those 
generally accepted for (Ga,Mn)As, the best understood 
(III,Mn)V ferromagnet.  $J_{pd}=0.1$eV nm$^3$,
$a_0=0.3$ nm, $\epsilon=10$, $V_0=2.5eV$ and $r_0=0.259$ nm.  
We diagonalize the HF Hamiltonian in a plane wave representation, using
periodic boundary conditions in the simulation cell.   
For different system sizes we maintain the cutoff wavevector $k_{max} =\pi$ nm$^{-1}$.
\end{references}
\end{document}